\documentclass[prb,nofootinbib,twocolumn,superscriptaddress]{revtex4} 

\usepackage{graphicx}
\usepackage{dcolumn}
\usepackage{bm}
\usepackage{threeparttable}
\usepackage{times}
\usepackage{mathptmx}
\usepackage{lscape}
\usepackage{natbib}
\usepackage{amsmath}
\usepackage{amssymb}
\usepackage{braket}

\def\degree{\kern-.2em\r{}\kern-.3em}

\begin{document}


\title{ Extension of Configurational Polyhedra to Finite Temperature Property}

\author{Koretaka Yuge}
\affiliation{
Department of Materials Science and Engineering,  Kyoto University, Sakyo, Kyoto 606-8501, Japan\\
}%

\author{Kazuya Kojima}
\affiliation{
Department of Materials Science and Engineering,  Kyoto University, Sakyo, Kyoto 606-8501, Japan\\
}%

\author{Kazuhito Takeuchi}
\affiliation{
Department of Materials Science and Engineering,  Kyoto University, Sakyo, Kyoto 606-8501, Japan\\
}%

\author{Tetsuya Taikei}
\affiliation{
Department of Materials Science and Engineering,  Kyoto University, Sakyo, Kyoto 606-8501, Japan\\
}%

\begin{abstract}
{  Configurational polyhedora (CP) is a hyperpolyhedra on multidimensional configuration space, whose vertex (and edges) corresponds to upper or lower limit value of correlation functions for all possible atomic configuration on given lattice. In classical systems where physical property including internal energy and elastic modulus can be a linear map for structures considered, it is known that atomic configuration having highest (or lowerst) physical quantity should always locate on one of the vertices at absolute zero temperature. The present study extend the idea of CP to finite-temperature property (especially, focusing on internal energy), and successfully provides demonstration of how temperature dependence of internal energy in equilibrium state for alloys can be characterized in terms of the spatial constraint on the system, and is interpreted in terms of the density of states for \textit{non-interacting} system along specially selected direction on cofiguration space.    }
\end{abstract}


\maketitle

\section{Introduction}
For alloys, when we want to obtain set of atomic configuration exhibiting extremal physical property (e.g., having lowest internal energy, having highest elastic modulus) for ground-state (i.e., at absolute zero temperature), we should in principle calculate their values over all possible atomic configuration on given lattice. Based on first-principles calculation, this direct approach should be practically intractable even for a limited  system size, since number of possible configuration takes $R^{N}$, where $R$ denotes number of components and $N$ denotes number of atom in the system. 
Therefore, alternative approaches has been typically applied to overcome such problem, where cluster expansion (CE) is one of the most well-established method to accurately predict configurational properties based on DFT calculations. CE employs Ising-like spin variables to specify atomic occupation on each lattice point, and corresponding effective cluster interactions (ECIs) can be obtained through DFT energy for a number of atomic configuration: Applying the ECIs to statistical thermodynamics simulation including Monte Carlo (MC) simulation, a set of atomic configuration having extremal physical property can be systematically predicted.

Although the combination of CE and DFT is powerful, a number of DFT calculation should still be required to obtain accurate ECIs. Use of configurational polyhedra (CP), which is hyper polyhedron in multidimensional cofiguration space, can be another approach: In classical systems, atomic configuration having maximum (or minimum) phisical quantity (including energy and elastic modulus) should always correspond to one of the vertices of CP. Since CP can be constructed without any information about energy, we can \textit{a priori} know a set of atomic configuration having extremal physical property for any combination of constituent elemtns. The CP depends only on the spatial constraint on the system: For instance, for crystalline solids, crystal lattice (e.g., fcc, bcc, hcp and diamond) corresponds to the constraint. 
This naturally indicate the fact that spatial constraint on the system have significant role to determine extremal physical properties at absolute zero temperature. 
For finite temperature, we recently develop a theory based on statistical thermodynamics, clarifing how the spatial constraint connect with macroscopic property in  equilibrium states: We find that macroscopic physical properties and their temperature dependence in disordered states can be well-characterized by a few specially-selected atomic configurations that are established only from information about their underlying lattice. 
This indicates that for disordered states in classical systems, we can \textit{a priori} know a set of special microscopic states on configuration space, which characterizes macroscopic property. 
We also confirm that predicted temperature dependence of physical property (such as internal energy) by our theory deviates from that by full thermodynamic simulation with decrease of temperature near order-disorder transition temperature, while tendency of positive or negative deviation is not well-understood in terms of the spatial constraint: We believe this tendency of deviation can be another important key to further understand the role of spatial constraint on equilibrium properties.

In the present study, we extend the cencept of CP (original at absolute zero temperature) to considering finite temperature property, which can qualitatively capture tendency of the above deviation. We will introduce projected density of microscopic states for \textit{non-interacting} system along specially selected direction on configuration space, and show how their landscape relates to temperature dependence of internal energy in equilibrium state. 

\section{Concept and methodology}
We first briefly explain the concept of CP, and then, show how the information of CP can be extended to capture finite temperature property. 

In classical systems, potential energy $E$ (or other physical quantities) for any microscopic state $d$ can be given by a linear combination of complete basis functions ($q_s$) for 
configuration space:
\begin{eqnarray}
\label{eq:ce}
U ^{\left(d\right)} = \sum _{s=1} ^{g} \Braket{U|q_s}q_s ^{\left(d\right)},
\end{eqnarray}
where $\Braket{\quad|\quad}$ denotes inner product on configuration space. Since $\Braket{U|q_i}$ is constant in terms of microscopic states on configuration space, it is clear that 
constant energy region can be represented by a $g$-dimensional hyperplane in configuration space, where its gradient corresponds to a set of inner product in Eq.~(\ref{eq:ce}). 
Therefore, atomic configuration that have maximum (or minimum) physical quantity under constant inner products should locate on the corresponding hyperplane whose distance from center of gravity for density of microscopic states on configuration space in Euclidean metric takes highest along the considered direction. 
Then we can construct convex hyperpolyhedra consisting of all possible hyperplanes involving atomic configurations satisfying the above condition: This hyperpolyhedra is called as "Configurational Polyhedra". From the above discussions, it is now clear that atomic configuration having highest (or lowerst) physical quantity should always locate on one of the vertices of the CP. 

In our theory previously developed, we clarify a set of special microscopic states determining equilibrium properties, based on the characteristics of density of microscopic states (DOS) on configuration space: We confirm that for a wide class of spacial constraint on the system, the DOS for \textit{non-interacting system} can be well-characterized by multidimensional gaussian when number of constituents in the system increases. 
Using the characteristics of DOS and Eq.~(\ref{eq:ce}), we derive that canonical average of internal energy (or other physical quantity) can be given by
\begin{eqnarray}
\label{eq:lsi}
E\left( T \right) \simeq \Braket{E}_1 - \beta  \Braket{E}_2, 
\end{eqnarray}
where $\beta ^{-1}=k_{\textrm{B}}T$, and $\Braket{E}_m$ denotes $m$-th order moment of DOS in terms of internal energy over possible atomic configuration. 
\begin{figure}
\begin{center}
\includegraphics[width=0.8\linewidth]{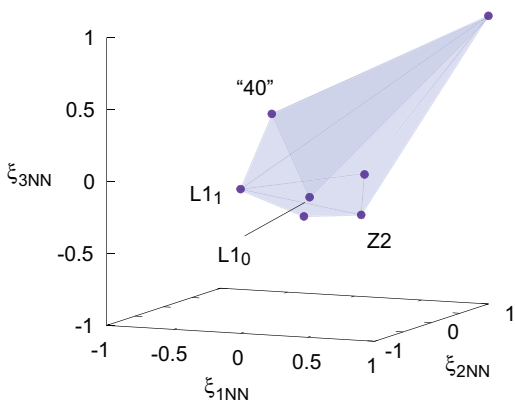}
\caption{Constructed configurational polyhedron for equiatomic fcc lattice, projected onto three dimensional configuration space in terms of 1NN, 2NN and 3NN pair correlation functions. }
\label{fig:cp3}
\end{center}
\end{figure}
It has been shown\cite{sqs} that $\Braket{E}_1$ can be estimated from a single special microscropic state, and we recently show\cite{lsi,emrs} that 
$\Braket{E}_2$ can also be estimated from a few special microscopic states established from geometrical characteristics of lattice. 
In Eq.~(\ref{eq:lsi}), left-hand exactly equals right-hand side when the DOS is completely represented by multidimensional gaussian,\cite{lsi} while the DOS for practical system far from its center of gravity should deviate from gaussian due to the existence of spatial constraint.\cite{emrs} 
Since Eq.~(\ref{eq:lsi}) is derived by using the fact that canonical average of physical quantity under the configurational DOS with multidimensional gaussian is invariant with respect to the system size,\cite{lsi,arxiv} the first-order correction to Eq.~(\ref{eq:lsi}) due to the deviation of DOS from gaussian should be $1/2\cdot \beta^{2}\Braket{E}_3$.
This certainly indicates that when the third-order moment of DOS in terms of energy is positive (negative), temperature dependence of internal energy above transition temperature tends to have positive (negative) deviation from Eq.~(\ref{eq:lsi}).  
Using this fact, we can extend the concept of CP to finite temperature property as follows: Since again, hyperplane in configuration space given by Eq.~(\ref{eq:ce}) with constant inner products corresponds to constant energy plane, we introduce projected DOS (PDOS) $g\left(\xi'_{v}\right)$ for each vertex of CP, $v$. $\xi'_{v}$ corresponds to a special coordination parallel to the vector from center of gravity of configurational DOS to the vertex $v$. 
Since hyperplane normal to $\xi'_{v}$ should always gives atomic configuration at vertex $v$ having extremal physical quantity, sign of the third-order moment of configurational DOS marginal to $\xi'_{v}$, i.e., $g\left(\xi'_{v}\right)$, characterizes the positive or negative deviation of temperature dependence of energy from Eq.~(\ref{eq:lsi}). The important point is that $g\left(\xi'_{v}\right)$ can be constructed for \textit{non-interacting} system, i.e., without information of energy or temperature. 

\section{Results and Discussions}
\begin{figure}[h]
\begin{center}
\includegraphics[width=0.93\linewidth]{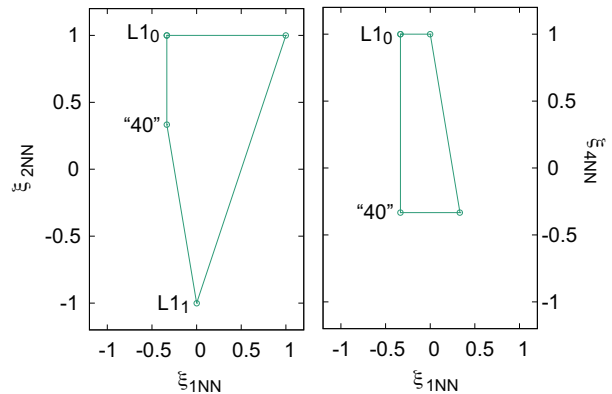}
\caption{Two dimensional landscape of configurational polyhedron in terms of 1NN and 2NN (left), and 1NN and 4NN (right) pair correlation functions.}
\label{fig:cp}
\end{center}
\end{figure}
Although the concept of CP is straightforwardly introduced, its concrete landscape for finite-size system is not completely understood even for representative lattice including fcc, bcc and hcp. 
In the present study, we first construct CP for equiatomic fcc lattice with a limited system size by calculating correlation functions for pairs up to 4-th nearest-neighbor (4NN) coordination for all possible atomic configurations within $2\times 2 \times 2$ expansion of convensional fcc unit cell, i.e., 32 atoms. We calculate correlation functions of up to 4NN pair for all possible atomic configurations (i.e., around 600 milion configurations), and the CP is constructed based on QuickHull algorism\cite{qh} finding a convex hull for given multiple points on configuration space.
Figure~\ref{fig:cp3} shows the constructed CP projected onto 3 dimensional space of 1NN, 2NN and 3NN pair correlation functions. 
We can find seven vertices, and the practical advantage of the CP can be easily demonstrated: Among the vertices, we can certainly confirm well-known ground-state ordered structures for practical binary alloys of L1$_0$ (for Cu-Au alloy), L1$_1$ for Cu-Pt, "40" for Pt-Rh. 
When we project the CP in terms of 1NN and 2NN pair correlation functions, its landscape becomes left-hand side of Fig.~\ref{fig:cp}, and for projection of 1NN and 4NN pair, it becomes right-hand side of Fig.~\ref{fig:cp}. We can clearly see from the figure that when we project the CP onto lower dimensional space, a set of vertices generally changes.

\begin{figure}[h]
\begin{center}
\includegraphics[width=0.8\linewidth]{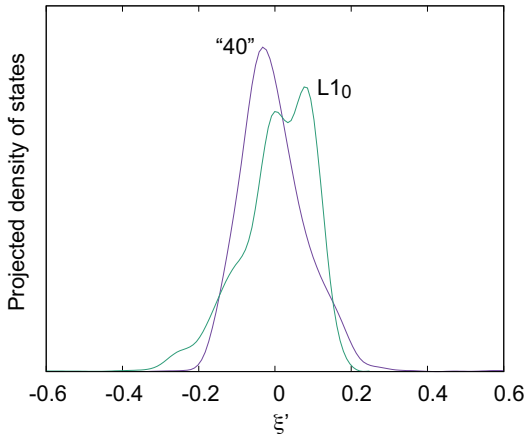}
\caption{Configurational density of microscopic states projected along specially selected coordination of $\xi'$. }
\label{fig:pdos}
\end{center}
\end{figure}

\begin{figure}[h]
\begin{center}
\includegraphics[width=0.95\linewidth]{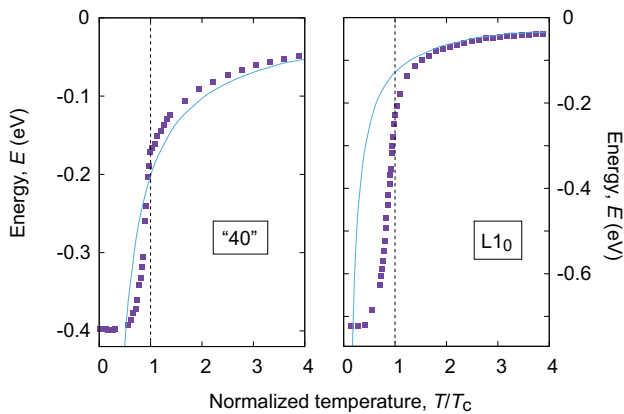}
\caption{Temperature dependence of configurational energy obtained from Monte Carlo simulation (closed squares) and from Eq.~(\ref{eq:lsi}) (solid curves), where the system in left-hand and righ-hand side has ground-state of ``40'' and L1$_{0}$, respectively.}
\label{fig:mc}
\end{center}
\end{figure}

Using the CP up to 4NN pair coordination, we can define a specially selected coordination $\xi'$ for each vertex, where in the present case, we consider two representative ordered structure of L1$_{0}$ and ``40'' as shown in Figs.~\ref{fig:cp3} and ~\ref{fig:cp}. 
Figure~\ref{fig:pdos} shows calculated PDOS along the special coordination of $\xi'$ for L1$_{0}$ and ``40''. We can clearly see that landscape of the PDOS strongly depends on the chosen coordination (or equivalently, chosen atomic configuration on vertex). Using the PDOSs, we estimate their third-order moment, $\mu_{3}$, measured from the center of gravity: $\mu_{3}$ for ``40'' takes positive value of 2.5$\times 10^{-4}$, while that for 
L1$_{0}$ takes negative value of -5.7$\times 10^{-4}$, which indicates that canonical average of configuraional energy for two binary systems individually have ground-state of ``40'' and L1$_{0}$ exhibit opposite temperature dependence in terms of Eq.~(\ref{eq:lsi}). 
In order to confirm the temperature dependence, we artificially prepare two set of ECIs up to 4NN pairs, where each set of ECIs results in ground-state of ``40'' and L1$_{0}$, respectively. These ECIs are applied to Monte Carlo statistical simulation under canonical ensemble to obtain temperature dependence of configurational energy. 
The results are shown in Fig.~\ref{fig:mc}, together with the analytical result using our theory of Eq.~(\ref{eq:lsi}). It is clear from 
Fig.~\ref{fig:mc} that temperature dependence of configurational energy for system ``40'' exhibits positive deviation from the theory, while that for 
L1$_{0}$ exhibits negative deviation. These positive and negative deviations are certainly consistent with the sign of the third-order moment for 
PDOS, $\mu_{3}$ as described above, which indicates that concept of the CP can be extended to characterize the temperature dependence of the physical quantity. 

\section*{Acknowledgement}
This work was supported by a Grant-in-Aid for Scientific Research C (16K06704) from the MEXT of Japan, Research Grant from Hitachi Metals$\cdot$Materials Science Foundation, and Advanced Low Carbon Technology Research and Development Program of the Japan Science and Technology Agency (JST).

\end{document}